\documentclass[twocolumn,showkeys,showpacs,preprintnumbers,superscriptaddress,nofootinbib]{revtex4}
\usepackage{graphicx}
\usepackage{dcolumn,amsmath,amssymb}
\usepackage{bm} 
\begin{document}


\title{\textsl{L.\,sibirica}~Ledeb. CHLOROPLAST GENOME YIELDS UNUSUAL SEVEN-CLUSTER STRUCTURE}

\author{Michael G.\,Sadovsky}
\affiliation{Institute of computational modelling of SD of RAS;\\ 660036 Russia, Krasnoyarsk, Akademgorodok.} \email{msad@icm.krasn.ru}
\author{Eugenia I.\,Bondar}
\affiliation{Siberian Federal university;\\ 660041 Russia, Krasnoyarsk, Svobodny prosp.79.} \email{bone-post@ya.ru}
\author{Yuliya A.\,Putintseva}
\affiliation{Siberian Federal university;\\ 660041 Russia, Krasnoyarsk, Svobodny prosp.79.} \affiliation{Institute of forest of SD RAS;\\ 660036 Russia, Krasnoyarsk, Akademgorodok.} \email{yaputintseva@mail.ru}
\author{Konstantin V.\,Krutovsky}
\affiliation{Georg-August-University of G\"{o}ttingen, B\"{u}sgenweg, 2, G\"{o}ttingen, D-37077, Germany}\email{konstantin.krutovsky@forst.uni-goettingen.de}
\affiliation{Siberian Federal university;\\ 660041 Russia, Krasnoyarsk, Svobodny prosp.79.}
\affiliation{N.\,I.\,Vavilov Institute of General Genetics of RAS; Gubkin Str., 3, Moscow, 119333, Russia}
\affiliation{Texas A\,\&\,M University, HFSB 305, 2138 TAMU, College Station, Texas, 77843, USA}

\begin{abstract}
We studied the structuredness in a chloroplast genome of Siberian larch. The clusters in 63-dimensional space were identified with elastic map technique, where the objects to be clusterized are the different fragments of the genome. A seven-cluster structure in the distribution of those fragments reported previously has been found. Unlike the previous results, we have found the drastically other composition of the clusters comprising the fragments extracted from coding and non-coding regions of the genome.
\end{abstract}


\maketitle

\section{Introduction}
Molecular biology provides mathematics with a number of mathematically sound problems and questions. Eventually, the structure identification and an order implementation in an ensemble of finite sequences are the most interesting among them. Finite symbol sequences, being a typical mathematical object, are naturally present as genetic matter in any living being; namely, as DNA sequence. Further we will consider the finite symbol sequence of chloroplast genomes of five plant species, including one from \textit{Larix sibirica}~Ledeb., which was recently completely sequenced, assembled and annotated in the Laboratory of Forest Genomics at the Genome Research and Education Centre of Siberian Federal University \cite{sfu1,bondar1}.

This sequence consisted of 122\,561 symbols or letters from the four-letter alphabet~$\aleph =\{\mathsf{A}, \mathsf{C}, \mathsf{G}, \mathsf{T}\}$. Neither other symbols, nor blank spaces are supposed to be found in a sequence; a sequence under consideration is also supposed to be coherent (i.\,e., consisting of a single piece).

An identification and search of structures in DNA sequence is a main objective of mathematical bioinformatics, biophysics and related scientific fields, including computer programming and information theory. Structures observed within a sequence reveal an order and provide easier understanding of functional roles of a sequence or its fragments. A new function (or a connection between function and structure, or taxonomy) might be discovered through a search for new patterns in symbol sequences corresponding to DNA molecule.

Previously, an intriguing seven-cluster structure in various genomes has been reported \cite{7k1,7k2,7k3}. In brief, the observed pattern consists of seven groups of considerably short fragments of a genome (say, $\sim 3\times 10^2$ nucleotides) arranged into seven clusters, in dependence on the information content encoded in those fragments. Three clusters comprising a triangle in the seven-cluster pattern gather the fragments encoding genes, etc., other three ones gather the fragments that are the complimentary ones to the former ones, and the seventh central cluster gathers the fragments to be found in the non-coding regions of a genome.

Here we report the similar structure observed over a chloroplast genome of \textit{L.\,sibirica}~Ledeb. Unlike the patterns described in \cite{7k1,7k2,7k3}, the structure observed in the chloroplast genome has drastically different pattern of the vertices to be found in the clusterization: the nodes comprising coding and non-coding regions are located in the basically different manner.

\section{Material and methods}
The chloroplast genome sequence Siberian larch (\textit{L.\,sibirica}~Ledeb.) has been sequenced using the Illumina HiSeq2000 sequencer at the Laboratory of Forest Genomics of the Siberian Federal University \cite{sfu1,bondar1} (see also \cite{sfu2}). The chloroplast genome contains 121 coding regions with 69 entities to be found in a leading strand, and 52 ones found in the lagging one, respectively. Total length of all the genes is equal to 68\,307~bp. The average length of a coding regions is 593~bp, ranging from  70~bp (the shortest one) to 6\,560~bp (the longest one). The standard deviation of the lengths of coding regions is $\sigma_{\mathsf{CDS}} = 1028.9$~bp. So great figure of the standard deviation results from a family of very short tRNA genes. The list of all genes found in the genome is provided in Table~\ref{tablogenov}.

\begin{table}
\caption{\label{tablogenov}\textit{Product} is a list of typical protein products of a group of genes; $M$ is the number of corresponding genes in the group}
\begin{tabular}{l|c}\hline
\multicolumn{1}{c|}{\textit{product}} & $M$\\\hline
tRNA & 31\\
rRNA & 3\\
Ribosomal proteins & 25\\
Photosystems I,II & 22\\
Cytochrome ($b_6{-}f$, $b_{559}$) & 10\\
RNA polymerase & 5\\
ATP synthase & 4\\
Light-independent protochlorophyllide reductase & 3\\
Translation initiation factor 1 & 1\\
NAD(P)H-quinone oxidoreductase & 1\\
Cell division protein FtsH & 1\\
Maturase K & 1\\
Ribulose bisphosphate carboxylase & 1\\
Acetyl-coenzyme A carboxyl transferase & 1\\
ATP-dependent Clp protease & 1\\
Hypothetical protein & 10\\\hline
\end{tabular}
\end{table}

Earlier \cite{osid98,sadov1,sadov2,sadov3} a frequency dictionary has been shown to be a fundamental structure of a symbol sequence. Consider a symbol sequence $\mathfrak{T}$ of the length~$N$ from the four-letter alphabet~$\aleph$ mentioned above. No gaps take place in a text. The word~$\omega = \nu_1\nu_2\ldots\nu_{q-1}\nu_q$ of the length~$q$ is a string occurred in the text~$\mathfrak{T}$. Here~$\nu_j$ is a symbol occupying the $j$-th position at the word; $\nu_j \in \aleph$.

Everywhere below we shall consider the 3 symbols (nucleotides) long words, only (and call them triplets). $(q,l)$-frequency dictionary~$W_q(l)$ is the set of all the words of the length~$q$ counted within the text~$\mathfrak{T}$ with the step in~$l$ symbols, so that each word is accompanied with its frequency. A frequency of a word~$\omega$ is defined traditionally: that is the number~$n_{\omega}$ of copies of the word divided by the total number of all copies of all the words \cite{osid98,sadov1,sadov2,sadov3}. Parameter~$l$ is arbitrary in a dictionary; everywhere further we will consider only the $W_3(3)$ frequency dictionaries. Evidently, a $W_3(3)$ frequency dictionary comprises a set of codons in some cases.

Triplets play the key role in inherited information processing, and this is the basic idea standing behind the choice. Besides, we follow the classic papers \cite{7k1,7k2,7k3} where the triplet frequencies $W_3$ have been used to develop the cluster structure of bacterial and yeast genomes. The authors of \cite{7k1,7k2,7k3} had also tried the other $q$-tipple combinations (for $q=2$ and $q=4$, respectively) and found that such choice provides significantly less information towards the structuredness of a nucleotide sequence. Another informal support for this choice comes from the observation over the information capacity of various genomes \cite{preprint}, where triplets also are shown to be featured from other $q$-tipples, with $q\neq 3$.

A frequency dictionary $W_3(3)$ unambiguously maps a text~$\mathfrak{T}$ into a 64-dimensional space with triplets being the coordinates, and the frequencies are the coordinate figures. Hence, frequency dictionary represents a short range (or meso-scale, at most) structuredness in a symbol sequence. Consider, then, a window selecting a fragment~$\mathfrak{F}$ of the length~$S$ in a text~$\mathfrak{T}$. Then $R(S,d)$ is an $(S,d)$-lattice that is the set of the fragments of the length~$S$ consequently selected alongside the text~$\mathfrak{T}$ by the window of the length~$S$, with the step~$d$. Obviously, a lattice consists of overlapping fragments, if~$d<S$.

That is the basic object for further analysis of statistical properties of a symbol sequence representing chloroplast genomes. The key idea of the paper is to check whether the fragments obtained for some $(S,d)$-lattice $R(S,d)$ differ in their statistical properties, or not. The properties expressed in the terms of $(3,3)$-frequency dictionaries $W_3(3)$ would be considered, only.

\subsection{Clusterization techniques}
We used the approach to figure out clusters in a dataset based on an elastic map technique \cite{g1,gusev,g2}. The basic idea of this method is to approximate the multidimensional data with a manifold of smaller dimension; the elastic map technique implies the approximation with two-dimensional manifold (see details in \cite{gusev}). In brief, the procedure looks like the following. At the first step, the first and the second principal components must be found. Then a plane must be developed over these two axes. At the second step, each data point must be projected at the plane and connected with the projection by an elastic spring. At the third step, the plane is allowed to bend and expand; so, the system is to be released to reach the minimum of the total energy (deformation plus spring extension). At the fourth step, each data point must be re-determined on the jammed map. Namely, a new data point image is the point on the map that is the closest to the original point in terms of the chosen metrics. Finally, the jammed map is ``smoothened'' by inverse non-linear transformation (for more details see \cite{gusev,g2}). All the results were obtained with \textsl{ViDaExpert} software by A.~Zinovyev\footnote{http://bioinfo-out.curie.fr/projects/vidaexpert/}.

\section{Results}
To begin with, we describe the procedure of the development of the data set. Firstly, the genome sequence was covered with the $(S,d)$-lattice; $S=303$, $d=10$. It was important that $d \neq 0\!\!\pmod 3$. Each fragment in the lattice was labeled with the number of central symbol of the given fragment. Next, each identified fragment of the lattice has been transformed into~$W_3(3)$ frequency dictionary. Hence, the sequence was mapped into a set of the points in a metric 64-dimensional space. We used Euclidean metrics hereafter.

Since the linear constraint
\begin{equation}\label{1}
\sum_{\omega} f_{\omega} = 1
\end{equation}
brings an additional parasitic signal, one of the triplets must be eliminated from the set. Formally, any triplet could be eliminated, but, practically, the choice may affect the results of the further treatment. Two strategies could be implemented here: either to exclude the triplet with the greatest frequency, or to remove the triplet yielding the least contribution into the points separation and discrimination. We persuaded the second strategy: the triplet~$\mathsf{CGC}$ with the minimal standard deviation $\sigma_{\mathsf{CGC}} = 0.00655$ over the entire dataset was excluded. Finally, the dataset of 12\,226 points arranged in 63-dimensional Euclidean space has been obtained; each point here corresponds to some fragment of the genome.

\begin{figure*}
\centerline{\includegraphics[width=9cm]{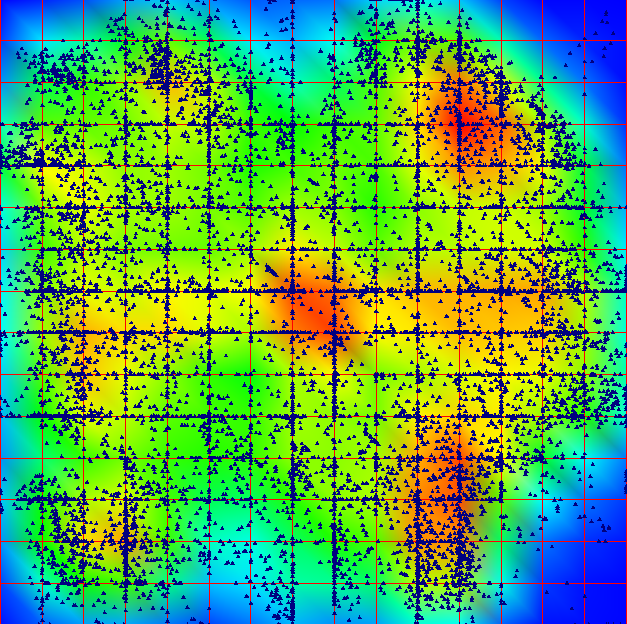}\hfill\includegraphics[width=9.1cm]{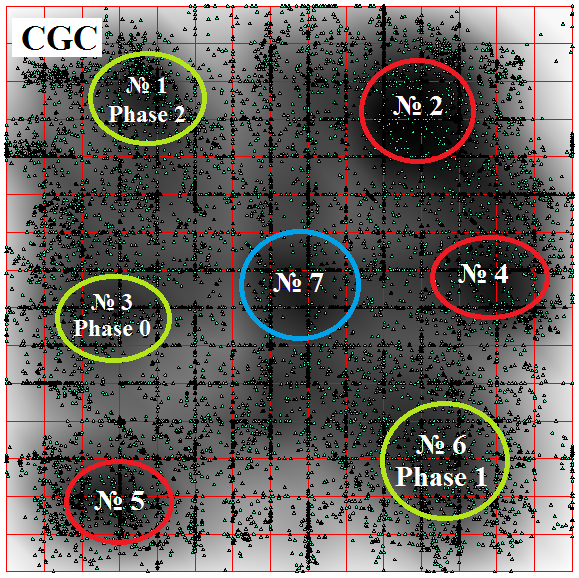}}
\caption{\label{chloro_unst}The seven-cluster structure identified over \textit{L.\,sibirica}~L. chloroplast genome (left). The right figure shows the distribution of the fragments from coding vs. non-coding regions of the genome: clusters \#~1, \#~3 and \#~6 comprise the fragments from coding regions; clusters \#~2, \#~4 and \#~5 gather the fragments from non-coding regions. Finally, the cluster encircled in blue (\#~7) present both coding and non-coding regions, almost equally in parts.}
\end{figure*}

To figure out the clusters over the dataset mentioned above, we explore the elastic map technique \cite{g1,gusev,g2}; \textsl{ViDaExpert} by A.\,Zinovyev has been used. The standard parameters configuration was used to develop the map (see Fig.~\ref{chloro_unst}, left) that depicts the famous seven cluster structure \cite{7k1,7k2,7k3} more explicitly with due edge-node pattern. The left picture in the figure shows the distribution of the fragments over the clusters; colors indicate a local density of the fragments, with maximal labeled in red, and the lowest one labeled in blue.

The right picture in Fig.~\ref{chloro_unst} shows the distribution of the fragments located within coding vs. non-coding regions of the genome. To begin with, we shall explain the idea of phase of a fragment. The (absolute) phase of a fragment labeled by the number $S_j$ (here $S_j$ denotes the position of the central nucleotide alongside the genome) yields three figures for the remainder of the division of $S_j$ by~3: 0, 1 and 2. These figures make the absolute phase of a fragment.

Meanwhile, an absolute phase may have nothing to do with a biological charge of a fragment: indeed, it measures the location of a fragment against the first nucleotide in a sequence, while a gene (or a coding region) may have, or may have not the fixed absolute phase. Thus, one has to introduce the relative phase determining the location of a fragment with respect to a coding region. The location (expressed in nucleotide numbers) of a gene or another functionally charged site to be found within the genome is provided by the genome annotation. Hence, we determined the relative location of each fragment against the coding region, whereas a fragment is embedded into the region. Again, the relative phase is defined as the reminder of the division by~3 of the length of the string connecting the start position of the coding region, and the fragment; obviously, the relative phase yields the figures of~0, 1 and 2. Hence, the relative phase identifies the fragments in $(S,d)$-lattice starting at the same position (by the reminder, to be exact) within any coding region.

Fig.~\ref{chloro_unst} (right picture) shows the distribution of the relative phases over the clusters obtained due to the procedure of clusterization described above (for $(303, 10)$-lattice). It should be stressed that the relative phase is defined for the fragments falling inside a coding region, only. Obviously, no phase is defined for the fragments located within the non-coding regions; simultaneously, we did not determine relative phases for the fragments occupying the cluster~\#~7 comprising a mixture of the fragments located both inside the coding region and outside these former.

\section{Discussion}
Seven-cluster structures in various genomes (bacterial, mainly) were reported earlier \cite{7k1,7k2,7k3}. It has been found that the pattern is provided by two triangles whose vertices are the clusters; whether one would observe seven-cluster or four-cluster structure, severely depends on $\mathsf{GC}$-content of a genome. For the genomes with the content close to an equilibrium one (say, about $0.40 \div 0.46$), the structuredness manifests through the seven-cluster pattern. A bias of the $\mathsf{GC}$-content to some of poles yields a kind of degeneration of the pattern into a four-cluster structure. The $\mathsf{GC}$-content for the chloroplast genome under consideration is equal to $0.43$ what makes the observed seven-cluster structure to be concordant to the previously reported ones.

Surely, the choice of the specific figures for $(S,d)$-lattice may affect strongly the clusterization results. Yet, we have no comprehensive and substantial idea how to choose the figures. Probably, some heuristic approaches could be used for that. For the technical reasons, $S$ must be odd and divisible by~3; reciprocally, $d$ may vary significantly, while it must not be divisible be~3. A growth of $d$ length results just in a decay of the capacity of the dataset: the total number of fragments to be taken into consideration goes down (linearly). A choice of $S$ figure is less evident: the figure right and above illustrates this fact. It shows the similar seven-cluster pattern observed for a rice chloroplast genome, with $S=3003$ (see Fig.~\ref{fig3}). Probably, a natural constraint to choose the $S$ figure is to take it close to (an average) length of a gene or other functionally charged site to be found within the genome. This idea makes the figure of $S=303$ taken for the studies described above rather natural and informative.

\begin{figure}
\includegraphics[width=8cm]{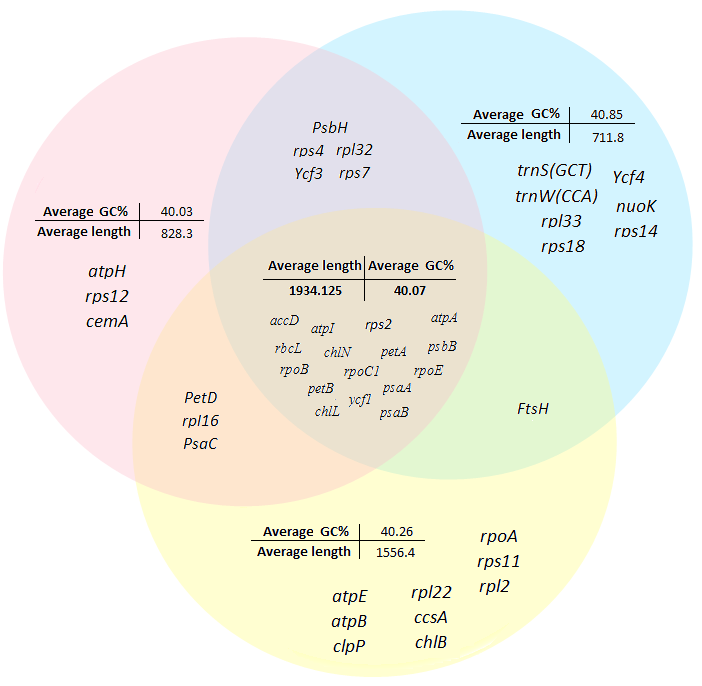}
\caption{\label{fig2}Distribution of genes over the  clusters.}
\end{figure}
An obvious bias in the mutual location of the fragments from coding vs. non-coding region raises a new question towards the distribution of genes between those clusters (these are the clusters \#~1, \#~3 and \#~6 in Fig.~\ref{chloro_unst}). Fig.~\ref{fig2} answers this question: evidently non-equilibrium distribution of the genes is observed over the clusters. The greater part of genes are irrelevant to the cluster occupation; moreover, the idea of a leading role of $\mathsf{GC}$-content in the distribution of genes over the clusters failed. Next group of genes (in terms of abundance of that latter) is found in a sole cluster. Finally, the least number of genes are distributed among two clusters (see Fig.~\ref{fig2}). The observed pattern is not a matter of surprise: this is one more manifest of the (very diverse) relation of a structure and a function. Yet, the pattern should be checked and verified: we did not distinguish the genes located in leading strand from those located in lagging one. Taking into account such disposition, one may expect that the pattern changes slightly. Definitely, it will not be destroyed completely, nor it would change into absolutely another form; nonetheless, some minor while important changes may take place.

A distribution of the fragments located within coding and non-coding regions of the genome makes a significant deviation from the results presented in \cite{7k1,7k2,7k3}. The authors of \cite{7k1,7k2,7k3} stipulate that the nodes of the seven-cluster structure (as shown in Fig.~\ref{chloro_unst}) could be arranged into two triangles, where the first triangle comprises the fragments from the coding regions, the second triangle also comprises the coding regions (while with dual triplets that make the so called complimentary palindromes, or the couples defined according to the Chargaff's parity rule), and the central cluster gathers the fragments located within the non-coding regions. This might be true for bacterial genomes which are known to bear very few non-coding regions.

Roughly, the ratio of the length of coding vs. non-coding regions for \textsl{L.\,sibirica} Ledeb. chloroplast genome is about~$0.5$. Maybe, this fact results in the pattern observed for the genome (see Fig.~\ref{chloro_unst}): two dual triangles comprise the fragments belonging to coding and non-coding regions, separately. An immediate and to some extent innocent explanation of this observation may come from the length of the genome: it might be short enough, so that the finite sampling effects take place manifesting through this separation of coding vs. non-coding fragments. Since chloroplast genomes are not (typically) too long, then one may address the question making simulation. Taking a bacterial genome (or better any other with close figures for $\mathsf{GC}$-content and coding vs. non-coding regions ratio), one can take several parts of such genome, randomly chosen, to be a surrogate ``chloroplast genome''. If similar pattern is observed, then the finite sampling effect takes place.

If no finite sampling effect takes place, then the observed pattern is expected to result from biological issues of the chloroplast genomes. To figure out whether the biology affects the combinatorial properties of the nucleotide sequences, one should carry out a comparative study of the clusterization over a family of chloroplast genomes of various species; this topic falls beyond the scope of the paper.

\begin{figure}
\includegraphics[bb=0 0 397 305, viewport=0 0 397 290, clip=true, scale=0.45]{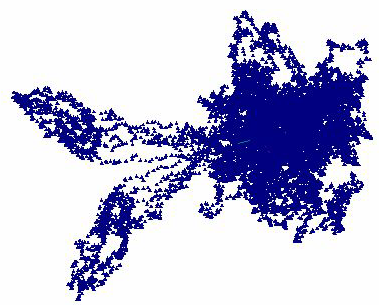}
\caption{\label{fig3}An example of clusterization similar to that one shown in Fig.~\ref{chloro_unst}, with window length $S=3000$. The pattern is shown in principal components visualization scheme.}
\end{figure}
Here we present some preliminary results on clusterization observed over the chloroplast genome of \textsl{L.\,sibirica} Ledeb. Originally, the seven-cluster structure has been observed for bacterial or yeast genomes \cite{7k1,7k2,7k3}. Chloroplast are supposed to take origin in bacterial world. Nonetheless, the structure we have found differs strongly, from a similar one observed on bacterial genomes. Thus, further studies should address the comparative analysis of the pattern described above with the similar studies of chloroplast genomes of \textit{Abies sibirica} and \textit{Pinus sibirica}; obviously, one should compare the patterns observed on conifers, with those to be observed over angiosperm species.

Additionally, we have collected a number of portraits of chloroplast genomes of some species; the collection supports the idea of a clusterization of coding vs. non-coding fragments, while the pattern may differ significantly form a ``classical'' seven-cluster one. Everywhere further in these pictures orange colored points correspond to non-coding fragments, and crimson indicates phase~0, green indicates phase~1,yellow indicates phase~2. Just enjoy the pictures!

\section{Conclusion}
Seven cluster structure in chloroplast genome for \textit{L.\,sibirica} was found. This is the fundamental structure of any genome; the found pattern is not degenerated since frequency of nucleotide $\mathsf{A}$ differed significantly from frequency of nucleotide $\mathsf{G}$. The absence of a degeneracy may indicate the prototypic genome that gave origin to chloroplasts entities; it is supposed to be a bacterial one (following symbiotic theory of organelle origin). Unlike nuclear genome of bacteria, the chloroplast genome yields more complex structure of (at least two) clusters: these seem to consists of two and three subclusters, respectively. The detailed structure of these complex clusters needs more studies, but may bring new understanding of a fine structure details, or of relations between structure and function of chloroplast genome.

\section{Acknowledgements}
This study was supported by a research grant No.~14.Y26.31.0004 from the Government of the Russian Federation. We also thank Maria Yu.\,Senashova from ICM SB RAS (Krasnoyarsk) for help in calculations and data visualization.

\newpage
\begin{figure*}
\includegraphics[width=18cm]{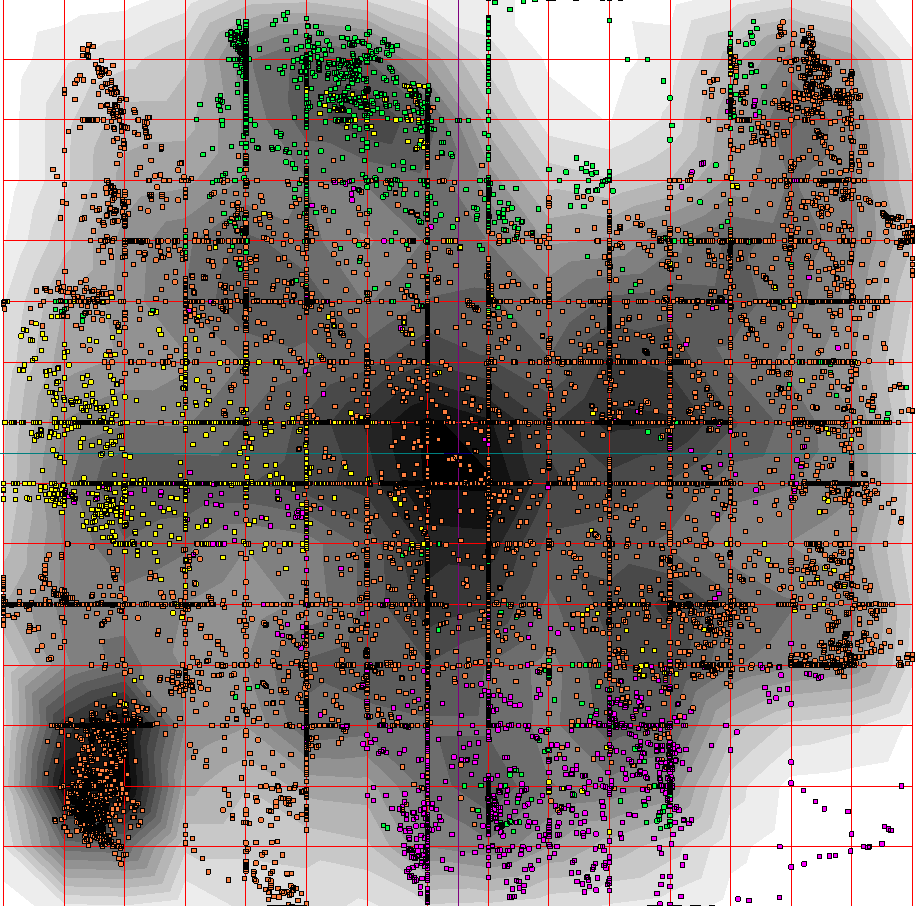}
\caption{\label{fdo1}\textit{Larix decidua} chloroplast DNA; accession number is AB501189.}
\end{figure*}

\begin{figure*}
\includegraphics[width=18cm]{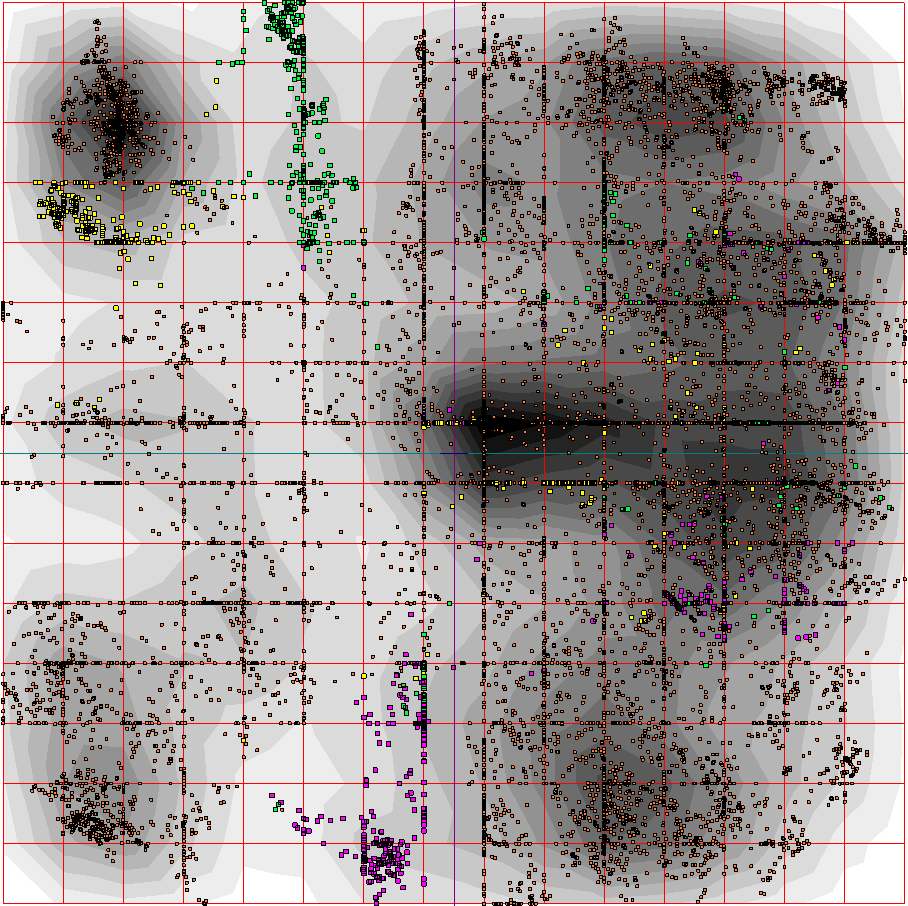}
\caption{\label{fdo2}\textit{Physcomitrella patens} chloroplast DNA; accession number is AP005672. That is moss.}
\end{figure*}

\begin{figure*}
\includegraphics[width=18cm]{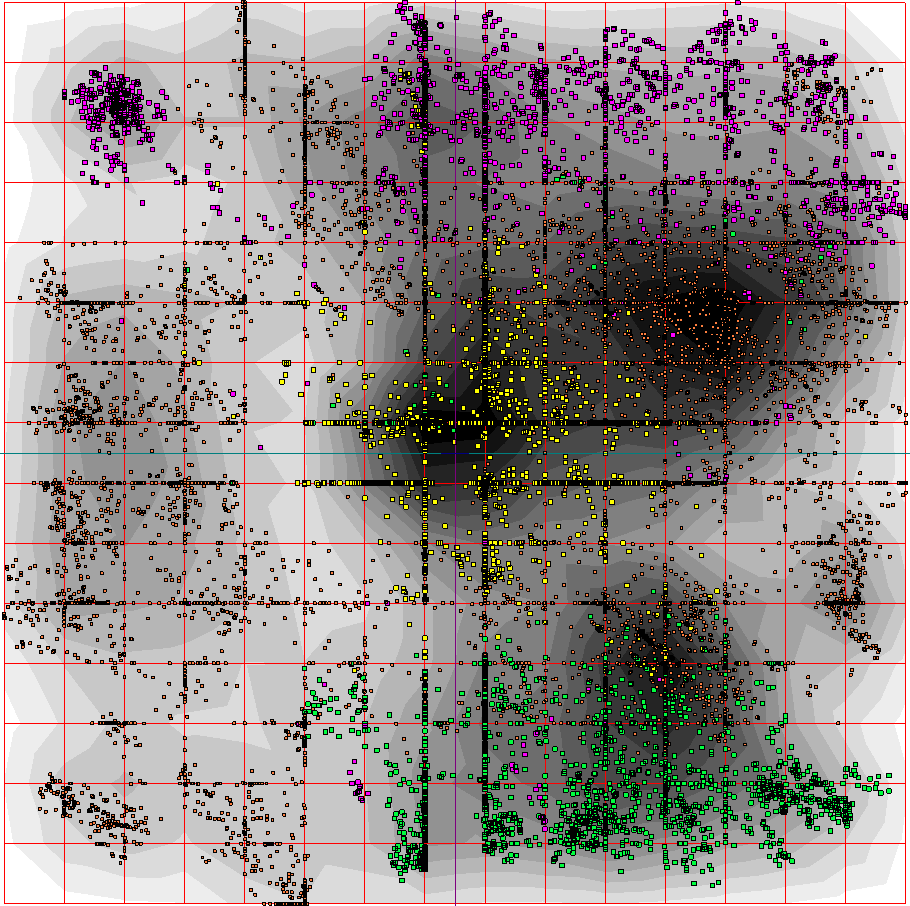}
\caption{\label{fdo3}\textit{Equisetum arvense} chloroplast DNA; accession number is GU191334. That is mare's-tail.}
\end{figure*}

\begin{figure*}
\includegraphics[width=18cm]{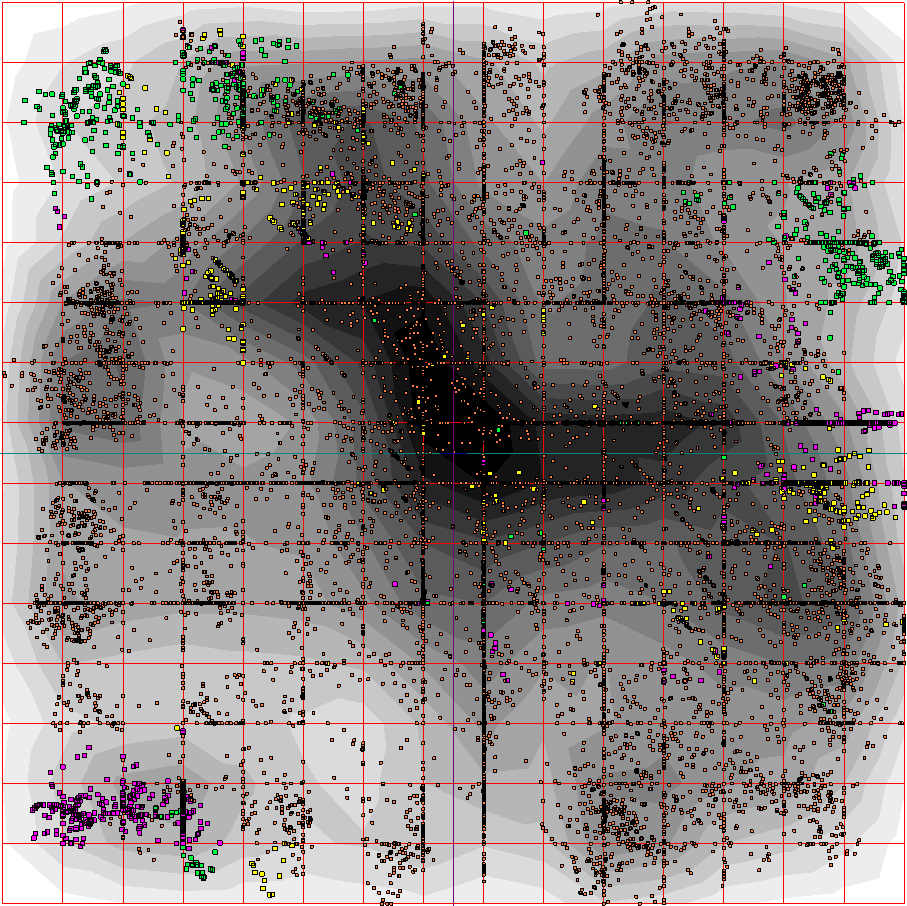}
\caption{\label{fdo4}\textit{Ginkgo biloba} chloroplast DNA; accession number is AB684440.}
\end{figure*}

\begin{figure*}
\includegraphics[width=18cm]{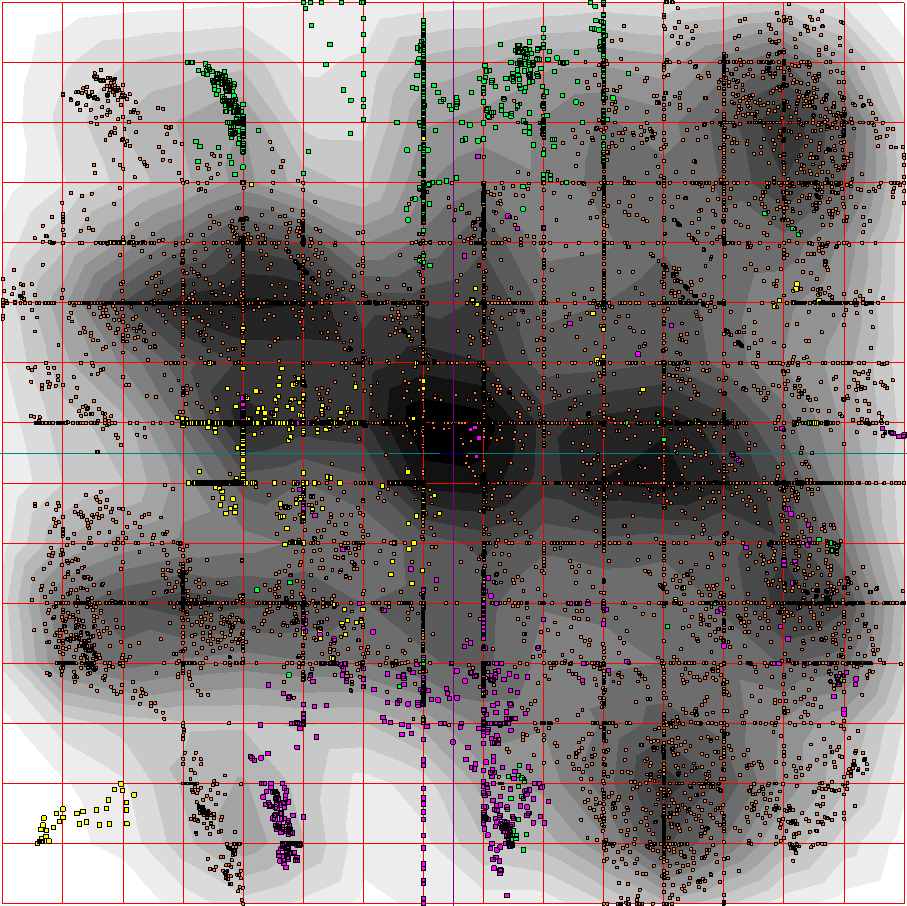}
\caption{\label{fdo5}\textit{Pinus taeda} chloroplast DNA; accession number is KC427273. That is pine.}
\end{figure*}

\begin{figure*}
\includegraphics[width=18cm]{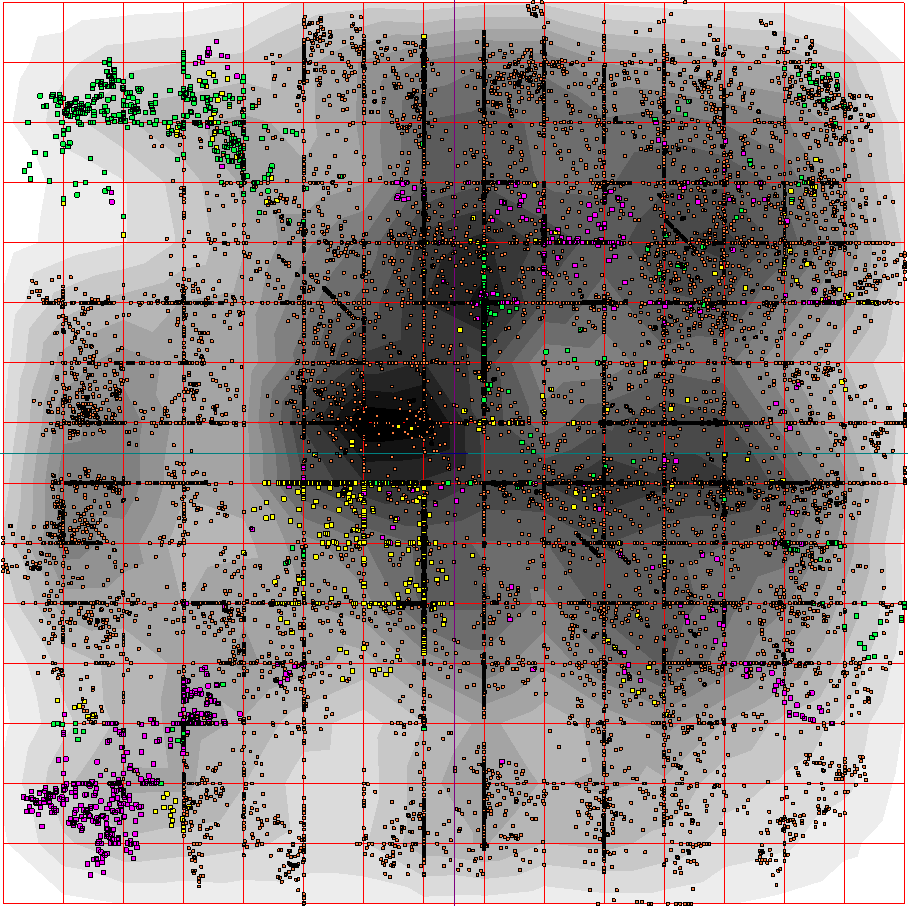}
\caption{\label{fdo6}\textit{Arabidopsis thaliana}  (thale cress) chloroplast DNA; accession number is AP000423.}
\end{figure*}

\begin{figure*}
\includegraphics[width=18cm]{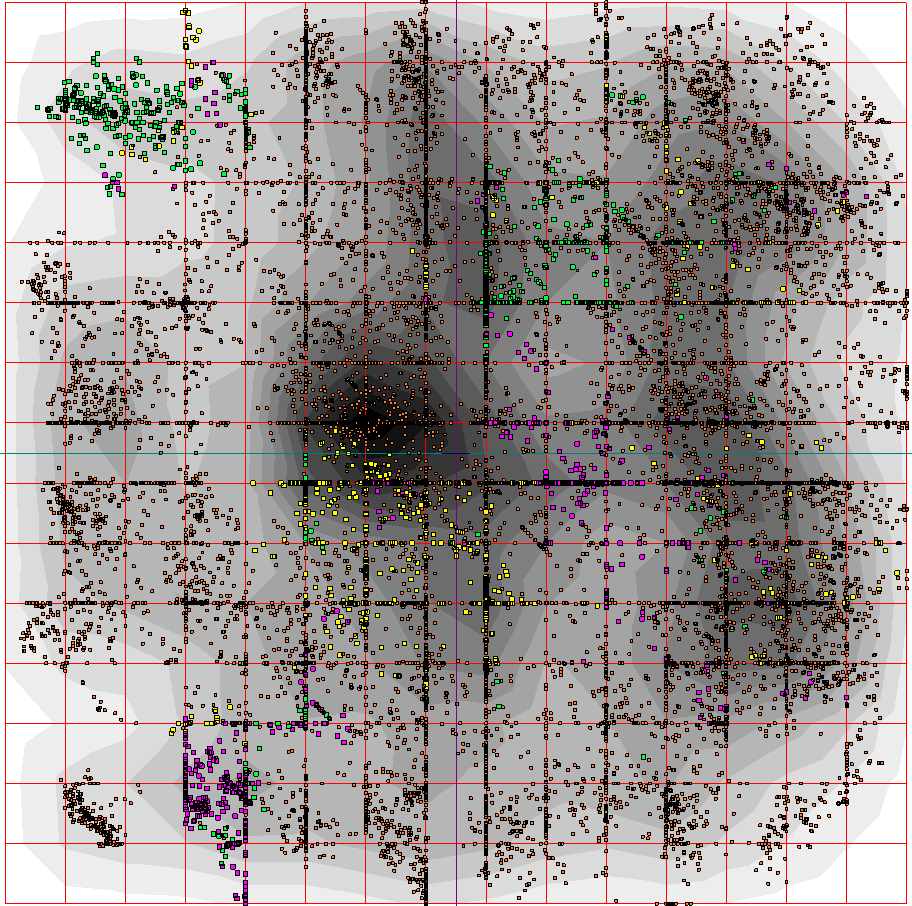}
\caption{\label{fdo7}\textit{Populus alba} chloroplast DNA; accession number is AP008956. That is silver poplar.}
\end{figure*}
\end{document}